\documentclass[12pt,a4paper,fleqn,twoside]{article}

\usepackage{amsmath}
\usepackage{amssymb}

\numberwithin{equation}{section} 
\newtheorem{theorem}{Theorem}[section]
\newtheorem{lemma}[theorem]{Lemma}
\newtheorem{definition}[theorem]{Definition}
\newtheorem{example}[theorem]{Example}

\newenvironment{proof*}{\paragraph{Proof.}}{}

\renewcommand{\pmatrix}[2]{\left ( \begin{array}{#1} #2 \end{array} \right )}

\newcommand{\qed}{\square}
\newcommand{\Dq}{\mathcal{D}}
\newcommand{\Lq}{\mathcal{L}}
\newcommand{\itbf}[1]{\textit{\textbf{#1}}}
\newcommand{\arrow}{\rightarrow}
\newcommand{\map}{\mapsto}

\newcommand{\bb}[1]{\mathbb{#1}}

\newcommand{\Diff}[3]{\left . \frac{d}{d#1}#2\right |_{#3}}

\newcommand{\Si}{\mathbb{S}^1}
\newcommand{\Cs}{\mathcal{C}}
\newcommand{\Ds}{\mathcal{D}}

\newcommand{\me}{\geqslant}
\newcommand{\les}{\leqslant}
\newcommand{\Dx}[1]{\partial_{x}^{#1}}
\newcommand{\Dy}[1]{\partial_{y}^{#1}}
\newcommand{\deriv}[2]{\frac{\partial #1}{\partial #2}}
\newcommand{\bra}[1]{\left (#1\right )}
\newcommand{\pobr}[1]{\left \{#1\right \}}
\newcommand{\Matrix}[3]{\left (
\begin{array}{c}
#2\\ #3
\end{array}
\right )_{#1}}
\newcommand{\Matrixx}[4]{\left (
\begin{array}{c}
#2\\ #3\\ #4
\end{array}
\right )_{#1}}
\newcommand{\var}[2]{\frac{\delta #1}{\delta #2}}

\newcommand{\ad}{{\rm ad}}
\newcommand{\tr}{{\rm tr}}
\newcommand{\res}{{\rm res}}
\newcommand{\Tr}{{\rm Tr}}

\begin{document}

\title{Classical $R$-matrix theory of dispersionless systems:\\II. (2+1)-dimension theory}

\author{Maciej B\l aszak\footnote{E-mail: blaszakm@amu.edu.pl}$\ $ and B\l a\.zej M. Szablikowski\footnote{E-mail: bszablik@amu.edu.pl }\\Institute of Physics, A.Mickiewicz University,\\Umultowska 85, 61-614 Pozna\'n, Poland}

\maketitle

\begin{abstract}
A systematic way of construction of (2+1)-dimensional dispersionless integrable Hamiltonian systems is presented. The method is based on the so-called central extension procedure and classical R-matrix applied to the Poisson algebras of formal Laurent series. Results are illustrated with the known and new (2+1)-dimensional dispersionless systems.\\(To appear in J. Phys. A: Math. Gen.)
\end{abstract}

\section{Introduction}

Dispersionless integrable Hamiltonian systems are often considered
as a quasi-classical limit of the related soliton systems
(Takasaki and Takebe \cite{TT}, Konopelchenko and Alonso \cite{KA}
and the literature quoted there).
Nevertheless, it seems that a more systematic approach, allowing a
construction of such systems from scratch, is necessary. Actually,
we are interested in a systematic way of construction of a class
of dispersionless systems having a Hamiltonian structure, and
infinite hierarchy of symmetries and conservation laws. One method
of doing it is based on the classical $R$-matrix theory. As well
known, the $R$-matrix formalism proved very fruitful in a
systematic construction of soliton systems (see for example
\cite{S-T-S}-\cite{Bl1} and the literature quoted there). So, it seems
reasonable to develop such a formalism for dispersionless systems.
Recently, an important progress in that direction was made by
Luen-Chau Li \cite{Li}. In paper \cite{Disp1} we apply his results
to a particular class of Poisson algebras \cite{G-KR} in order to construct
multi-Hamiltonian (1+1)-dimensional dispersionless systems.

Having such an effective theory for constructing multi-Hamiltonian
dispersionless dynamical systems in (1+1)-dimensions, we were
prompted to extend this method onto (2+1)-dimensions. The central
extension was considered in early works by Reyman and
Semenov-Tian-Shansky \cite{RS-T-S1,RS-T-S2} and also by
Prykarpatsky \cite{PSA1,PSA2}. The central extension approach to
integrable field and lattice-field systems was presented as well
in \cite{BS,BSP}.

As our construction leads in general to nonlocal equations, we
will understand by a dispersionless systems in (2+1)-dimension
PDEs of the form
\begin{equation}
\frac{\partial u_{i}}{\partial
t}=\sum_{j=1}^{n}v_{ij}(u,\Dq)\frac{\partial u_{j}}{\partial x}
+\sum_{j=1}^{n}w_{ij}(u,\Dq)\frac{\partial u_{j}}{\partial y}
,\qquad i=1,...,n,  \label{1}
\end{equation}
where $v_{ij}$ and $w_{ij}$ are pseudo-differential operators of
formal symbols $\Dq \equiv \Dx{-1} \Dy{}$.

The paper is organized as follows. In section 2, we briefly present
a number of basic facts and definitions of Hamiltonian dynamics on
Poisson algebras concerning the formalism applied. In section 3, we
present the general formulation of the central extension procedure
on Poisson algebras. In section 4 and 5, we apply this and the
$R$-matrix procedure to the Poisson algebras of formal Laurent
series. Then in section 6 we illustrate our results with the known
and new integrable Hamiltonian (2+1)-dimensional dispersionless
dynamical systems.

\section{Hamiltonian dynamics on Poisson algebras:\\$R$-structures}

Here, we repeat some basic facts presented in Part I to make the
paper selfconsistent. The reader familiar with Part I may skip
this section.

\begin{definition}
Let $A$ be a commutative, associative algebra with unit $1$. If
there is a Lie bracket on $A$ such that for each element $a\in A$,
the operator $\ad_a:b\map [a,b]$ is a derivation of the
multiplication, then $(A,[.,.])$ is called a \itbf{Poisson
algebra}.
\end{definition}
Thus, the Poisson algebras are Lie algebras with an additional
associative algebra structure (with commutative multiplication and
unit $1$) related by the derivation property to the Lie bracket.

Let $A$ be a Poisson algebra, $A^*$ the dual algebra related to
$A$ by the {\it duality map} $\langle \cdot,\cdot \rangle \arrow
\bb{R}$,
\begin{equation}
A^* \times A \arrow \bb{R}:\quad (\alpha,a)\map \langle \alpha,a
\rangle ,\qquad a\in A,\ \alpha \in A^*,
\end{equation}
and $\Ds(A^*):=\bb{C}^\infty(A^*)$ be a space of
$\bb{C}^\infty$-functions on $A^*$.  Let $F\in \Ds(A^*)$, then a
map $dF:A \arrow A$ such that
\begin{equation}\label{grad}
\Diff{t}{F(L+tL')}{t=0} = \langle L',dF(L) \rangle,\qquad L,L'\in
A^*,
\end{equation}
is a gradient of $F$.

We confine our further considerations to such Poisson algebras $A$
for which its dual $A^*$ can be identified with $A$. So, we assume
the existence of a product $(\cdot,\cdot)_A$ on $A$ which is
symmetric, non-degenerate and $\ad$-invariant:
\begin{equation}
(\ad_a b,c)_A + (b,\ad_a c)_A = 0, \qquad a,b,c\in A.
\end{equation}
Then, we can identify $A^*$ with  $A$, $(A^*\cong A)$ by setting
\begin{equation}
\langle \alpha,b \rangle = (a,b)_A,\qquad a,b\in A,\ \alpha \in
A^*,
\end{equation}
where $\alpha \in A^*$ is identified with $a\in A$.

\begin{definition}
A linear map $R:A \arrow A$ is called a {\bf classical $R$-matrix} if the R-bracket
\begin{equation}\label{lieR}
[a,b]_R:= [Ra,b]+[a,Rb],\qquad a,b\in A,
\end{equation}
is a second Lie product on $A$.
\end{definition}

\begin{theorem} \cite{Li}
Let $A$ be a Poisson algebra with Lie bracket $[\cdot,\cdot]$ and
non-degenerate ad-invariant pairing $(\cdot,\cdot)_A$ with respect
to which the operation of multiplication is symmetric, i.e.
$(ab,c)_A=(a,bc)_A,\ \forall a,b,c\in A$. Assume $R\in {\rm End}(A)$ is
a classical $R$-matrix, then for each integer $n\me -1$, the
formula
\begin{equation}\label{pobr}
\{H,F\}_n = (L,[R(L^{n+1}dF),dH]+[dF,R(L^{n+1}dH)])_A,
\end{equation}
where $H,F$ are smooth functions on $A$, defines a Poisson
structure on $A$. Moreover, all $\{\cdot,\cdot\}_n$ are
compatible.
\end{theorem}

The related Poisson bivectors $\pi_n$ are given by the following
Poisson maps
\begin{equation}\label{pob}
\pi_n :dH\map -\ad_L R(L^{n+1}dH) - L^{n+1} R^* (\ad_L dH),\qquad
n\me -1,
\end{equation}
where the adjoint of $R$ is defined by the relation
\begin{equation}
(a,Rb)_A=(R^*a,b)_A.
\end{equation}

Notice that the bracket \eqref{pobr} with $n=-1$ is just a
Lie-Poisson bracket with respect to a Lie bracket \eqref{lieR}
\begin{equation}\label{liepor}
\{H,F\}_{-1} = (L,[dF,dH]_R)_A.
\end{equation}

We will look for a natural set of functions in involution w.r.t.
the Poisson brackets \eqref{pobr}. A smooth function $F$ on $A$ is
$ad$-invariant if $dF\in \ker ad_L$, i.e
\begin{equation}
[dF,L]=0,\qquad L\in A,
\end{equation}
which are Casimir functionals of the natural Lie-Poisson bracket.

Hence, the following Lemma is valid
\begin{lemma} \cite{Li}
 Smooth functions on $A$ which are ad-invariant commute in $\{\cdot,\cdot\}_n$. The Hamiltonian system generated by a smooth ad-invariant function $C(L)$ and the Poisson structure $\{\cdot,\cdot\}_n$ is given by the Lax equation
\begin{equation}\label{eveqt}
L_t = [R(L^{n+1}dC),L],\qquad L\in A.
\end{equation}
\end{lemma}

For any $R$-matrix each two evolution equations in the hierarchy
\eqref{eveqt} commute due to the involutivity of the Casimir
functions $C_q$. Each equation admits all the Casimir functions as
a set of conserved quantities in involution. In this sense we will
regard \eqref{eveqt} as a hierarchy of \emph{integrable} evolution
equations.

Let us assume that an appropriate product on Poisson algebra $A$
is given by {\it the trace form} $\tr:A \arrow \mathbb{R}$
\begin{equation}\label{trA}
(a,b)_A = \tr(ab),\qquad a,b\in A.
\end{equation}

To construct the simplest $R$-structure let us assume that the
Poisson algebra $A$ can be split into a direct sum of Lie
subalgebras $A_+$ and $A_-$, i.e.
\begin{equation}
A =A_+ \oplus A_-,\qquad [A_\pm,A_\pm]\subset A_\pm.
\end{equation}
Denoting the projections onto these subalgebras by $P_\pm$, we
define $R$-matrix as
\begin{equation}\label{rp}
R = \frac{1}{2} (P_+ - P_-)
\end{equation}
which is well defined.

Two following Lemmas \cite{O},\cite{Bl1} are useful in calculating
of Hamiltonians $H(L)$ from the gradients $dH(L)$
\begin{lemma}
(Poincare). If $M$ is a linear space, or more generally is of the
star shape $(\forall_{L\in M} \{ \lambda L:0\les \lambda \les 1 \}
\subset M)$, each closed $k$-form is exact.
\end{lemma}
\begin{lemma}\label{fun}
Let $M$ fulfill the condition of the Poincare Lemma. Then for an
exact one-form $\gamma(L)$
\begin{equation}
H(L) = \int_0^1 \langle \gamma(\lambda L),L \rangle\ d\lambda
\end{equation}
is a zero-form such that $dH(L) = \gamma(L)$.
\end{lemma}

 Following the above scheme, we are able to construct in a systematic way integrable multi-Hamiltonian dispersionless systems, with infinite hierarchy of involutive constants of motion and infinite hierarchy of related commuting symmetries, ones we fix a Poisson algebra.

\section{Central extension approach}

Assume now that the Poisson algebra $A$ depends effectively on an
independent parameter $y\in \Si$, which naturally generates the
corresponding current operator algebra $\Cs(A) =
\Cs^{\infty}(\Si,A)$ with the following modified $\Tr$-operation:
\begin{equation}
\Tr(a) := \int_{\Si} \tr(a)dy,
\end{equation}
where $\tr$ \eqref{trA} operation is defined for the Poisson
algebra $A$. The scalar product reads
\begin{equation}
(a,b)_{\Cs(A)}:= \Tr(a\cdot b)
\end{equation}
for $a$ and $b\in \Cs(A)$. The current Poisson algebra $\Cs(A)$
can be naturally extended via the central extension procedure:
$\Cs(A)\arrow \overline{\Cs}(A)=\Cs(A)\otimes \bb{C}$ with the
following Lie product:
\begin{equation}\label{lie_ext}
[(a,\alpha),(b,\beta)]:=([a,b],\omega_{2}(a,b)),
\end{equation}
where $\alpha,\beta \in \bb{C}$ and $\omega_{2}:\Cs(A)\times
\Cs(A)\arrow \bb{C}$ is the standard Maurer-Cartan two-cocycle on
$\Cs(A)$:
\begin{equation}
\omega_{2}(a,b):= \int_{\Si} \bra{a,\deriv{b}{y}}_A dy = \Tr(a\cdot
b_{y}),\qquad a,b\in \Cs(A).
\end{equation}
Recall that the Maurer-Cartan two-cocycle on a Lie algebra is a
bilinear $\bb{C}$-valued function satisfying two conditions:
\begin{enumerate}
\item[(i)] it is skew-symmetric
\begin{equation}
\omega_{2}(a,b) = -\omega_{2}(b,a);
\end{equation}
\item[(ii)] it satisfies the Jacobi identity
\begin{equation}
\omega_{2}([a,b],c)+\omega_{2}([c,a],b)+\omega_{2}([b,c],a)=0.
\end{equation}
\end{enumerate}
Hence, the Lie product (\ref{lie_ext}) is well defined on
$\overline{\Cs}(A)$. The scalar product on $\overline{\Cs}(A)$ is
given by
\begin{equation}\label{sc_ext}
((a,\alpha),(b,\beta))_{\overline{\Cs}(A)} := \Tr(a\cdot b) +
\alpha \cdot \beta.
\end{equation}

The Poisson bracket $\{\cdot,\cdot\}$ on the functionals
$\Ds(\Cs(A))$ we define as
\begin{align}\label{po_ext}
 \{H,F\}(L)&:= ((L,1),[(dF,1),(dH,1)])_{\overline{\Cs}(A)}\nonumber\\
&\; =((L,1),([dF,dH],\omega_{2}(dF,dH)))_{\overline{\Cs}(A)},
\end{align}
for all $(L,1)\in \overline{\Cs}(A^{*})\cong \overline{\Cs}(A)$.
Then from \eqref{sc_ext} we get the following form
\begin{equation}\label{exliepo}
\{H,F\}(L)= (L,[dF,dH])_{\Cs(A)} + \omega_{2}(dF,dH),
\end{equation}
which can be considered as a centrally extended Lie-Poisson
bracket.

Let us repeat the $R$-matrix approach for the current Lie
algebra $\overline{\Cs}(A)$ with a natural Lie-Poisson bracket
\eqref{exliepo}.
\begin{lemma}
Casimir functionals $C\in \Ds(\Cs(A))$ of a Lie-Poisson bracket
\eqref{exliepo} satisfy the so-called Novikov-Lax equation
\begin{equation}\label{novlax}
[dC,L] + (dC)_y = 0,
\end{equation}
for all $L\in \Cs(A^{*})\cong \Cs(A)$.
\end{lemma}
\begin{proof*}
For every $H,F\in \Ds(\Cs(A))$ and $L\in \Cs(A)$
\begin{align*}
\{H,F\}(L)&= (L,[dF,dH])_{\Cs(A)} + \omega_{2}(dF,dH)\\
 &= (dF,[dH,L])_{\Cs(A)} + (dF,(dH)_y])_{\Cs(A)} = (dF,[dH,L] +
(dH)_y])_{\Cs(A)},
\end{align*}
hence for Casimir functionals $C\in \Ds(\Cs(A))$
\begin{equation*}
\{C,F\}(L) = 0 \iff [dC,L] + (dC)_y = 0.\qquad \qed
\end{equation*}
\end{proof*}
The $R$-structure $\overline{R} \in {\rm End}(\overline{\Cs}(A))$
is defined as follows:
\begin{equation}
[(a,\alpha),(b,\beta)]_{\overline{R}}:=([a,b]_{R},\omega^{R}_{2}(a,b)),
\end{equation}
where $\omega^{R}_{2}(a,b):=\omega_{2}(R a,b)+\omega_{2}(a,R
b)$. Then, the new linear Lie-Poisson bracket has the following
form
\begin{align}\label{po_lin}
\{H,F\}_1 (L)&=
((L,1),[(dF,1),(dH,1)]_{\overline{R}})_{\overline{\Cs}(A)}\nonumber\\
&=(L,[dF,dH]_{R})_{\Cs(A)} + \omega^{R}_{2}(dF,dH).
\end{align}
\begin{lemma}
The following Poisson operator is related to the linear Poisson
bracket (\ref{po_lin}):
\begin{equation}\label{linP}
\theta(L):dH\map -ad_{L}R dH-R^{*}ad_{L}dH+(R
dH)_y+R^{*}(dH)_y .
\end{equation}
\end{lemma}
\begin{proof*}
For every $H,F\in \Ds(\Cs(A))$ and $L\in \Cs(A)$
\begin{align*}
\{H,F\}_{1}(L)&= (L,[dF,dH]_{R})_{\Cs(A)} + \omega^{R}_{2}(dF,dH)\\
&= (R dF,[dH,L])_{\Cs(A)} + (dF,[R dH,L])_{\Cs(A)} + (R dF,(dH)_y)_{\Cs(A)}\\
&\quad + (dF,(R dH)_y)_{\Cs(A)}\\
&= (dF,-[L,R dH]-R^{*}[L,dH]+(R dH)_y +R^{*}
(dH)_y)_{\Cs(A)}\\
&= (dF,\theta(L)dH)_{\Cs(A)}.\qquad \qed
\end{align*}
\end{proof*}
\begin{theorem}
The Casimir functionals $C_{i}\in \Ds(\Cs(A))$ of the Poisson
bracket (\ref{po_ext}) on $\Cs(A^{*})\cong \Cs(A)$ are in
involution with respect to the linear Poisson bracket
(\ref{po_lin}). Moreover, Casimir functionals $C_i$ satisfy the
following hierarchy of evolution equations:
\begin{equation}\label{eveq}
L_{t_i} = \theta(L)dC_{i} = [R dC_{i},L]+(R dC_i)_y,\qquad
i\in \bb{Z}.
\end{equation}
\end{theorem}
\begin{proof*}
Let $C_i$ and $C_j\in \Ds(\Cs(A))$ are Casimir functionals, then
\begin{align*}
&\{C_{i},C_{j}\}_{1}(L)= (L,[dC_{j},dC_{i}]_{R})_{\Cs(A)} + \omega^{R}_{2}(dC_{j},dC_{i})\\
&= (R dC_{j},[dC_{i},L]+(dC_{i})_y)_{\Cs(A)}+(R
dC_{i},[L,dC_{j}]-(dC_{j})_y)_{\Cs(A)} = 0.
\end{align*}
The proof of the second part of the theorem is obvious.$\qquad \qed$
\end{proof*}

In a special case of Poisson algebras, which are considered in the
paper, the bracket \eqref{po_lin} is nothing else but a centrally
extended Lie-Poisson bracket \eqref{liepor}. For higher order
Poisson brackets \eqref{pobr} we failed to prove the Poisson
property (Jacoby identity) after central extension.

\section{Poisson algebras of formal Laurent series}

 Let $A$ be an algebra of Laurent series with respect to $p$
\begin{equation}
A =\left \{L = \sum_{i\in \bb{Z}} u_i(x)p^i\right \},
\end{equation}
where the coefficients $u_i(x)$ are smooth functions. It is
obviously commutative and associative algebra under
multiplication. The Lie-bracket can be introduced in infinitely
many ways as
\begin{equation}\label{liebr}
[L_1,L_2]= p^r (\deriv{L_1}{p} \deriv{L_2}{x} - \deriv{L_1}{x}
\deriv{L_2}{p}):=\{L_1,L_2\}_r,\qquad r\in \bb{Z},
\end{equation}
as $\ad_L = p^r (\deriv{L}{p} \deriv{}{x} - \deriv{L}{x}
\deriv{}{p})$ is a derivation of the multiplication, so
$A_r:=(A,\{\cdot,\cdot\}_r)$ are Poisson algebras. An appropriate
symmetric product on $A_r$ is given by a trace form
$(a,b)_A:=\tr(ab)$:
\begin{equation}\label{trres}
\tr L = \int_\Omega \res_r Ldx,\qquad \res_r L = u_{r-1}(x),
\end{equation}
which is $\ad$-invariant. In expression \eqref{trres} the integration
denotes the equivalence class of differential expressions modulo
total derivatives. For a given functional $F(L)= \int_\Omega
f(u)dx$, we define its gradient as
\begin{equation}\label{res}
dF = \var{F}{L} = \sum_i \var{f}{u_i}p^{r-1-i},
\end{equation}
where $\delta f/\delta u_i$ is a variational derivative.

We construct the simplest $R$-matrix, through a decomposition of
$A$ into a direct sum of Lie subalgebras. For a fixed $r$ let
\begin{equation}
\begin{split}
A_{\me -r+k}&= P_{\me -r+k}A   =\left \{L = \sum_{i\me -r+k} u_i(x)p^i\right \},\\
A_{< -r+k}&= P_{< -r+k}A   =\left \{L = \sum_{i< -r+k}
u_i(x)p^i\right \},
\end{split}
\end{equation}
where $P$ are appropriate projections. As we presented in
\cite{Disp1}, $A_{\me -r+k}, A_{< -r+k}$ are Lie subalgebras in
the following cases:
\begin{enumerate}
\item[1.] $\quad k=0,\ r=0$,
\item[2.] $\quad k=1,2,\ r\in \bb{Z}$,
\end{enumerate}
which one can see through a simple inspection. Then, the
$R$-matrix is given by the projections
\begin{equation}\label{rmat}
R = \frac{1}{2}(P_{\me -r+k} - P_{< -r+k}) = P_{\me -r+k} -
\frac{1}{2} = \frac{1}{2} - P_{< -r+k}.
\end{equation}
To find $R^*$ one has to find $P_{\me -r+k}^*$ and $P_{< -r+k}^*$
given by the orthogonality relations
\begin{equation}
(P_{\me -r+k}^*,P_{< -r+k}) = (P_{< -r+k}^*,P_{\me -r+k})=0.
\end{equation}
So, we have
\begin{equation}
P_{\me -r+k}^* = P_{< 2r-k},\qquad P_{< -r+k}^* = P_{\me 2r-k},
\end{equation}
and then
\begin{equation}
R^* = \frac{1}{2}(P_{\me -r+k}^* - P_{< -r+k}^*) = \frac{1}{2} -
P_{\me 2r-k} = P_{< 2r-k} - \frac{1}{2}.
\end{equation}

\section{Centrally extended Poisson algebras of Laurent series}

Let $A$ be an algebra of Laurent series with respect to $p$
\begin{equation}
A =\left \{L = \sum_{i\in \bb{Z}} u_i(x,y)p^i\right \},
\end{equation}
where the coefficients $u_i(x,y)$ are smooth functions of two
variables $x$ and $y$. As in (1+1)-dimensional case $p$ was a
conjugate coordinate related to $x$, let us now introduce $q$ as a
conjugate coordinate related to $y$. Then, introducing the
extended Lie-bracket \eqref{liebr} in the form
\begin{equation}
\pobr{L_1,L_2}_r := p^r (\deriv{L_1}{p} \deriv{L_2}{x} -
\deriv{L_1}{x} \deriv{L_2}{p}) + \deriv{L_1}{q} \deriv{L_2}{y} -
\deriv{L_1}{y} \deriv{L_2}{q},\quad r\in \bb{Z},
\end{equation}
and the extended Lax element $\Lq \equiv L-q$, $L\in A$. The
Lax-Novikov equation \eqref{novlax} takes the form
\begin{equation}\label{caseq}
\{dC,\Lq \}_r = 0,
\end{equation}
and the hierarchy of evolution equations \eqref{eveq} for Casimir
functionals $C(L)$ with $R$-matrix given by \eqref{rmat} has the
form of two equivalent representations
\begin{equation}\label{laxh}
\Lq_{t_i} = \{(dC_i)_{\me -r+k},\Lq \}_r = -\{(dC_i)_{< -r+k},\Lq
\}_r,\qquad i\in \bb{Z},
\end{equation}
which are Lax hierarchies.

To construct dispersionless (2+1)-dimensional integrable
equations, at first we have to solve equation \eqref{caseq}, which
can be done by putting
\begin{equation}\label{sol1}
dC_i = \sum_{j\les i} a_j p^j,\qquad i\me -r+k,
\end{equation}
or by
\begin{equation}\label{sol2}
dC_i = \sum_{j\me i} a_j p^j,\qquad i< -r+k,
\end{equation}
where the function parameters $a_j$ are obtained from
\eqref{caseq} successively via the recurrent procedure. Notice
that although the solutions \eqref{sol1} or \eqref{sol2} are in
the form of infinite series, in fact we need only their finite
parts $(dC_i)_{\me -r+k}$ or $(dC_i)_{< -r+k}$. Hence, for a given
$\Lq$ in principle we can construct two different hierarchies of
Lax equations \eqref{laxh}.

We have to explain what type of Lax operator can be used in
(\ref{laxh}) to obtain a consistent operator evolution equivalent
to some nonlinear integrable equation. Obviously, we are
interested in extracting closed systems for a finite number of
fields. Hence, we start with looking for Lax operators $\Lq$ in
the general form
\begin{equation}\label{laxo}
\Lq = u_N p^N + u_{N-1} p^{N-1} + ... + u_{-m+1} p^{-m+1} + u_{-m}
p^{-m} - q
\end{equation}
of N-th order, parametrized by finite number of fields $u_i$. To
obtain a consistent Lax equation, the Lax operator (\ref{laxo})
has to form proper submanifold of the full Poisson algebra under
consideration, i.e. the left and right-hand sides of expression
(\ref{laxh}) have to lie inside of this submanifold.

Observing (\ref{laxh}) with some $(dC)_{< -r+k}=
a_{-r+k-1}p^{-r+k-1}+a_{-r+k-2}p^{-r+k-2}+...$ one immediately
obtains the highest order of the right-hand side of Lax equation
as
\begin{align}\label{high}
\Lq_{t} &= (u_N)_t p^N + (u_{N-1})_t p^{N-1} + ...\nonumber\\
 &= -\{(dC)_{< -r+k},u_N p^N + lower\}_r - \Dy{}(dC)_{< -r+k}\nonumber\\
 &= \bra{-((-r+k-1)a_{-r+k-1}(u_N)_x - N(a_{-r+k-1})_x u_N) p^{N+k-2}+ lower}\nonumber\\
 &\quad + \bra{- (a_{-r+k-1})_y p^{-r+k-1} + lower},
\end{align}
where $lower$ represents lower orders. Observing (\ref{laxh}) with
some $(dC)_{\me -r+k}= ... +
a_{-r+k+1}p^{-r+k+1}+a_{-r+k}p^{-r+k}$ one immediately obtains the
lowest order of the right-hand side of Lax equation (\ref{laxh})
as
\begin{align}\label{low}
\Lq_{t} &= ... + (u_{-m+1})_t p^{-m+1} + (u_{-m})_t p^{-m}\nonumber\\
 &= \{(dC)_{\me -r+k}, higher + u_{-m} p^{-m}\}_r + \Dy{}(dC)_{\me -r+k}\nonumber\\
 &= \bra{higher +((-r+k)a_{-r+k}(u_{-m})_x - (-m)(a_{-r+k})_x u_{-m}) p^{-m+k-1}}\nonumber\\
 &\quad + \bra{higher + (a_{-r+k})_y p^{-r+k}},
\end{align}
where $higher$ represents higher orders. Simple consideration of
\eqref{high} and \eqref{low} with condition $N\me -m$ leads to the
admissible Lax polynomials with a finite number of field
coordinates, which form proper submanifolds of Poisson
subalgebras. They are given in the form
\begin{align}
 &k=0,\ r=0 :\nonumber\\\label{laxk0}
 &\quad \Lq = c_N p^N + c_{N-1} p^{N-1} + u_{N-2} p^{N-2} + ... + u_0 -q \qquad \mbox{for}\quad N\me 1,\\
 &k=1, r\in \bb{Z} :\nonumber\\\label{laxk1a}
 &\quad \Lq = c_N p^N +
u_{N-1} p^{N-1} + ... + u_{-m} p^{-m}-q \qquad \mbox{for}\quad N\me 1-r\me -m,\\ \label{laxk1b}
 &\quad \Lq = u_{-r} p^{-r} + u_{-r-1} p^{-r-1} + ... + u_{-m} p^{-m}-q \qquad \mbox{for}\quad -r\me -m,\\
 &k=2,\ r\in \bb{Z} :\nonumber\\\label{laxk2a}
 &\quad  \Lq = u_N p^N + ... + u_{1-m} p^{1-m} + c_{-m} p^{-m}-q \qquad \mbox{for}\quad N\me 1-r\me -m,\\\label{laxk2b}
 &\quad \Lq = u_N p^N + ... + u_{3-r} p^{3-r} + u_{2-r} p^{2-r}-q \qquad \mbox{for}\quad N\me 2-r,
\end{align}
where the $u_i$ are dynamical fields and $c_N,c_{N-1},c_{-m}$ are
arbitrary time independent functions of $x$ and $y$. Lax operators
for $k=0,1,2$: \eqref{laxk0},\eqref{laxk1a},\eqref{laxk2a}  form a
proper submanifold in (1+1)-dimension \cite{Disp1}, hence the Lax
dynamics induced by them can be reduced onto (1+1)-dimensional
space. Lax operators for $k=1$: \eqref{laxk1b} and $k=2$:
\eqref{laxk2b} do not form s proper submanifold in
(1+1)-dimension, hence the Lax dynamic induced by them is purely
(2+1)-dimensional effect, and they cannot be reduced onto
(1+1)-dimensional space.

Hence, we know the restricted Lax operators $\Lq$ we can now
investigate the form of gradients of Casimir functionals $dC_i$
given by \eqref{sol1} or by \eqref{sol2} which satisfy equation
\eqref{caseq}, as well as we can investigate some further simplest
admissible reductions of Lax operators.

\paragraph{The case of $k=0$.}
Let us consider Lax operators of the form \eqref{laxk0}. Then
observing \eqref{caseq} with some $dC_i = a_i p^i + a_{i-1}
p^{i-1} + lower$ one immediately obtains the conditions for the
highest terms of $dC_i$, since
\begin{multline}
\{a_i p^i + a_{i-1} p^{i-1} + a_{i-2} p^{i-2} + lower,\Lq \}_0 = -N(a_i)_x c_N p^{i+N-1}\\
 - (N c_N (a_i)_x + (N-1) c_{N-1}
(a_{i-1})_x) p^{i+N-2}+ lower = 0.
\end{multline}
Therefore $(a_i)_x=(a_{i-1})_x=0,\
ia_i(u_{N-2})_x-Nc_N(a_{i-2})_x=0$ and so on, hence \eqref{sol1}
has the following form
\begin{equation}\label{gck0}
dC_i = \alpha_i p^i + \alpha_{i-1} p^{i-1} +
\frac{i\alpha_i}{Nc_N}u_{N-2}p^{i-2} + a_{i-3} p^{i-3}  +
lower,\qquad i\me 0,
\end{equation}
where $\alpha_i,\alpha_{i-1}$ are arbitrary $x$ independent
functions. Observing \eqref{caseq} with $dC_i = higher + a_{i+1}
p^{i+1} + a_i p^i$ one obtains the conditions for the lowest terms
of $dC_i$, since
\begin{multline}
 \{higher + a_{i+1} p^{i+1} + a_i p^i,\Lq \}_0 =\\
 higher + ((i+1)a_{i+1}(u_0)_x+ia_i(u_1)_x-(a_i)_xu_1+(a_i)_y)p^i +
ia_i(u_0)_xp^{i-1} = 0.
\end{multline}
Accordingly $a_i=0$ and $a_{i-1}=a_{i-2}=...=0$ since $a_j$
depends linearly on $a_{j+1},a_{j+2},...,a_i$. Hence for $k=0$
there is only one Lax hierarchy for the $dC_i$ of the form
\eqref{gck0}. There are no any obvious further reductions of
$\Lq$.

\paragraph{The case of $k=1$.}
For Lax operators of the form \eqref{laxk1a} by observing
\eqref{caseq}, $dC_i$ given by \eqref{sol1} or \eqref{sol2} have
the following forms
\begin{align}\label{gck1a1}
&dC_i = \alpha_i p^i + \frac{i\alpha_i}{Nc_N}u_{N-1}p^{i-1} +
a_{i-2} p^{i-2}  + lower,\qquad &i\me -r+1,\\ \label{gck1a2}
&dC_i =
higher + a_{i+2} p^{i+2} + a_{i+1} p^{i+1} +
\alpha_i(u_{-m})^{-\frac{i}{m}}p^i,\qquad & i< -r+1,
\end{align}
where $\alpha_i$ is an arbitrary $x$-independent function. For Lax
operators of the form \eqref{laxk1b} by observing \eqref{caseq},
$dC_i$ given by \eqref{sol1} or \eqref{sol2} have the following
forms
\begin{align}\label{gck1b1}
& dC_i = \beta_i p^i -
\Dy{-1}(i\beta_i(u_{-r})_x+r(\beta_i)_xu_{-r})p^{i-1} + a_{i-2}
p^{i-2}  + lower, \qquad i\me -r+1,\\ \label{gck1b2}
& dC_i = higher +
a_{i+2} p^{i+2} + a_{i+1} p^{i+1} +
\alpha_i(u_{-m})^{-\frac{i}{m}}p^i,\qquad i< -r+1,
\end{align}
where $\alpha_i$ and $\beta_i$ are arbitrary $x$- and
$y$-independent functions, respectively.

We remark that, if $-m<1-r$ in $\Lq$, there is a further
admissible reduction of the equations \eqref{laxh}, given by
$u_{-m}=0$, since such reduced Lax polynomials still are of the
form \eqref{laxk1a} or \eqref{laxk1b}. We have to look for the
form of gradients of Casimir functionals after such a  reduction.
It is easy to see that by this reduction $u_{-m}=0$, the gradients
of Casimir functionals \eqref{gck1a1} and \eqref{gck1b1} preserve
the order of the highest terms, and the form. For gradients of
Casimir functionals \eqref{gck1a2} and \eqref{gck1b2} by this
reduction the lowest order disappear, and as all other terms
depend linearly on it, such gradients reduce to zero, except the
one case $(dC_i)_{<-r+1}=(L)_{<-r+1}$ which produces equation
$\Lq_{t_i}=-\Lq_y$. We can continue the reductions by putting
$u_{1-m}=0$, if the reduced $\Lq$ are still of the form
\eqref{laxk1a} or \eqref{laxk1b} and so on. Therefore, the
reductions are proper in general only for the gradients of Casimir
functionals in the form \eqref{gck1a1} and \eqref{gck1b1}.

\paragraph{The case of $k=2$.}
For Lax operators of the form \eqref{laxk2a} by observing
\eqref{caseq}, $dC_i$ given by \eqref{sol1} or \eqref{sol2} have
the following form
\begin{align}\label{gck2a1}
& dC_i = \alpha_i(u_{N})^{\frac{i}{N}}p^i + a_{i-1} p^{i-1} +
a_{i-2} p^{i-2} + lower,\qquad i\me -r+2,\\ \label{gck2a2}
& dC_i =
higher + a_{i+2} p^{i+2}- \frac{i\alpha_i}{mc_{-m}}u_{1-m}p^{i+1}
+ \alpha_i p^i,\qquad i< -r+2,
\end{align}
where $\alpha_i$ is arbitrary $x$-independent function. For Lax
operators of the form \eqref{laxk2b} by observing \eqref{caseq},
$dC_i$ given by \eqref{sol1} or \eqref{sol2} have the following
form
\begin{align}\label{gck2b1}
& dC_i = \alpha_i(u_{N})^{\frac{i}{N}}p^i + a_{i-1} p^{i-1} +
a_{i-2} p^{i-2} + lower,\qquad i\me -r+2,\\ \label{gck2b2}
& dC_i =
higher -
\Dy{-1}(i\beta_i(u_{2-r})_x-(2-r)(\beta_i)_xu_{2-r})p^{i+1} +
\beta_i p^i,\qquad i< -r+2,
\end{align}
where $\alpha_i$ and $\beta_i$ are arbitrary $x$- and
$y$-independent functions, respectively.

If $N>1-r$ in $\Lq$, there is a further admissible reduction of
equations \eqref{laxh}, given by $u_{N}=0$ since such reduced Lax
polynomials still are of the form \eqref{laxk2a} or
\eqref{laxk2b}. By analogous considerations as for $k=1$, these
reductions are proper in general only for the gradients of Casimir
functionals in the form \eqref{gck2a2} and \eqref{gck2b2}.

The different schemes are interrelated as it is explained in the
following theorem.
\begin{theorem}\label{rel}
Under the transformation
\begin{equation}\label{transf}
x'=x,\ y'=-y,\ p'=p^{-1},\ q'=q,\ t'=t
\end{equation}
the Lax hierarchy defined by $k=1,\ r$ and $\Lq$ transforms into the
Lax hierarchy defined by $k=2,\ r'=2-r$ and $\Lq'$, i.e.
\begin{equation}
k=1,\ r,\ \Lq \Longleftrightarrow k=2,\ r'=2-r,\ \Lq'.
\end{equation}
\end{theorem}
\begin{proof*}
It is readily seen that the Lax operators for $k=1$ and $r$ of the
forms \eqref{laxk1a}, \eqref{laxk1b} transform into the well
restricted Lax operators for $k=2$ and $r'=2-r$ of the forms
\eqref{laxk2a}, \eqref{laxk2b} respectively. Let's observe that
\begin{align*}
\{A,B\}_r &= p^r (\deriv{A}{p} \deriv{B}{x} - \deriv{A}{x} \deriv{B}{p}) + \deriv{A}{q} \deriv{B}{y} - \deriv{A}{y} \deriv{B}{q}\\
 &= -p'^{-r+2} (\deriv{A'}{p'} \deriv{B'}{x'} - \deriv{A'}{x'}
\deriv{B'}{p'}) - \deriv{A'}{q'} \deriv{B'}{y'} + \deriv{A'}{y'}
\deriv{B'}{q'} = -\{A',B'\}'_{r'}.
\end{align*}
and
\[ (dC)_{\me s}' = (dC')_{\les -s}.\]
Hence, we have
\begin{align*}
\Lq_t &= \{(dC)_{\me -r+1},\Lq\}_r = -\{(dC)'_{\me -r+1},\Lq'\}'_{r'}\\
 & =  -\{(dC')_{\les r-1},\Lq'\}'_{r'} =  -\{(dC')_{<
-r'+2},\Lq'\}'_{r'} = \Lq'_{t'}.\qquad \qed
\end{align*}
\end{proof*}
Therefore, some dispersionless systems can be reconstructed from
different Poisson algebras. Moreover, we remark that the
gradients of Casimir functionals for $k=1$ and $k=2$ transform by
$p^{-1}=p'$ reciprocally at slant, i.e.
\eqref{gck1a1}$\leftrightarrow$\eqref{gck2a2},
\eqref{gck1a2}$\leftrightarrow$\eqref{gck2b1} and
\eqref{gck1b1}$\leftrightarrow$\eqref{gck2b2},
\eqref{gck1b2}$\leftrightarrow$\eqref{gck2b1}.

Two equivalent representations of Poisson structure coming from
the linear Poisson tensor \eqref{linP} with the $R$-matrix given
by \eqref{rmat} are
\begin{align}\label{linten}
\theta(\Lq)dH &= \{(dH)_{\me -r+k},\Lq\}_r - (\{dH,\Lq\}_r)_{\me 2r-k}\nonumber\\
&= -\{(dH)_{< -r+k},\Lq\}_r + (\{dH,\Lq\}_r)_{< 2r-k}.
\end{align}
It turns out that the first representation yields a direct access
to the lowest polynomial order of $\theta dH$, whereas the second
representation yields the information about the highest orders
present. There are two options. The best situation is when a given
Lax operator forms a  proper submanifold of the full Poisson
algebra, i.e. the image of the Poisson operator $\theta$ lie in
the space tangent to this submanifold for each element. If this is
not the case, {\it the Dirac reduction} can be invoked for
restriction of a given Poisson tensor to a suitable submanifold.

\paragraph{The case of $k=0$.}
Let us first consider the simplest admissible Lax polynomial
\eqref{laxk0} of the form
\begin{equation}\label{lk0}
\Lq = p^N + u_{N-2} p^{N-2} + ... + u_1 p + u_0 - q.
\end{equation}
This is the well-known dispersionless Gelfand-Dickey case. Then,
the gradient of the functional $H(\Lq)$ is given in the form
\begin{equation}\label{dhk0}
\var{H}{\Lq} = \var{H}{u_0} p^{-1} + \var{H}{u_1} p^{-2} + ... +
\var{H}{u_{N-2}} p^{1-N}.
\end{equation}
By inserting \eqref{lk0} into \eqref{linten} it becomes clear from
the first representation of the linear tensor that lowest order of
$\theta dH$ is at least zero, from the second representation it is
evident that the highest differential order will be at most $N-2$.
Hence, $\theta dH$ is tangent to the submanifold formed by the Lax
operator of the form \eqref{lk0}. As a result, these Lax operators
form a proper submanifold of full Poisson algebra, and the Poisson
tensor, since $\bra{\var{H}{\Lq}}_{\me 0} = 0$, is given by
\begin{equation}\label{tenk0}
\theta \bra{\var{H}{\Lq}} = \bra{\pobr{\Lq,\var{H}{\Lq}}_0}_{\me 0}.
\end{equation}

\paragraph{The case of $k=1$.}
Let us first consider the simplest admissible Lax operator
\eqref{laxk1a} in the form
\begin{equation}\label{lk1a}
\Lq =  p^N + u_{N-1} p^{N-1} + ... + u_{1-m} p^{1-m} + u_{-m}
p^{-m} - q.
\end{equation}
Then gradient of functional $H(\Lq)$ is given in the form
\begin{equation}\label{dhk1}
\var{H}{\Lq} = \var{H}{u_{-m}} p^{r+m-1} + \var{H}{u_{-m+1}}
p^{r+m-2} + ... + \var{H}{u_{N-1}} p^{r-N}.
\end{equation}
Inserting \eqref{lk1a} into \eqref{linten} one immediately obtains
the highest and lowest order of $\theta dH$ as
\begin{align}
\theta dH &= \bra{ (...)p^{N-1} + lower} + \bra{ (...)p^{2r-2} + lower}\nonumber\\
&= \bra{ higher + (...)p^{-m} } + \bra{ higher + (...)p^{2r-1} },
\end{align}
where $lower$ ($higher$) represents lower (higher) orders. Hence,
Lax operators of the form \eqref{lk1a} form a proper submanifold
for $N\me 2r-1\me -m$, as then $\theta dH$ is tangent to this
submanifold. So the linear Poisson map is
\begin{equation}\label{tenk1a}
\theta \bra{\var{H}{\Lq}} = \pobr{\bra{\var{H}{\Lq}}_{\me
-r+1},\Lq}_r+\bra{\pobr{\Lq,\var{H}{\Lq}}_r}_{\me 2r-1}.
\end{equation}
Otherwise a Dirac reduction is required.

Analogously, for Lax operators \eqref{laxk1b} in the form
\begin{equation}\label{lk1b}
\Lq = u_{-r} p^{-r} + u_{-r-1} p^{-r-1} + ... + u_{1-m} p^{1-m} +
u_{-m} p^{-m} - q
\end{equation}
we have
\begin{align}
\theta dH &= \bra{ (...)p^{-r} + lower } + \bra{ (...)p^{2r-2} + lower }\nonumber\\
&= \bra{ higher + (...)p^{-m} } +\bra{ higher + (...)p^{2r-1} }.
\end{align}
Hence, this operator forms a proper submanifold for $r\les 0$ and
$2r-1\me -m$. The Poisson tensor is given by \eqref{tenk1a}. In
other cases a Dirac reduction is required. The simplest case is
$r=1$ with one-field reduction. Let
\begin{equation}
\overline{\Lq} = u + \Lq = u + u_{-1}p^{-1} + u_{-2}p^{-2} + ... +
u_{1-m}p^{1-m} + u_{-m}p^{-m} - q.
\end{equation}
The Dirac reduction with the constraint $u=0$ leads to the Poisson
map in the form
\begin{equation}\label{tenk1b}
\theta^{red}\bra{\var{H}{\Lq}} =
\bra{\pobr{\var{H}{\Lq},\Lq}_1}_{<1} +
\pobr{\Dy{-1}res_1\pobr{\Lq,\var{H}{\Lq}}_1,\Lq}_1,
\end{equation}
which is generally nonlocal.

\paragraph{The case of $k=2$.}
Let us consider Lax polynomials \eqref{laxk2a} in the form
\begin{equation}\label{lk2a}
\Lq = u_N p^N + u_{N-1} p^{N-1} + ... + u_{1-m} p^{1-m} +
p^{-m}-q.
\end{equation}
Then gradient of functional $H(\Lq)$ is given in the form
\begin{equation}\label{dhk2}
\var{H}{\Lq} = \var{H}{u_{1-m}} p^{r+m-2} + ... + \var{H}{u_{N-1}}
p^{r-N} + \var{H}{u_{N}} p^{r-N-1}.
\end{equation}
Then by analogous considerations as for $k=1$ or by Theorem
\ref{rel}, $\Lq$ given by \eqref{lk2a} form a proper submanifold
for $N\me 2r-3\me -m$. The Poisson tensor has the form
\begin{equation}\label{tenk2a}
\theta \bra{\var{H}{\Lq}} = \pobr{\bra{\var{H}{\Lq}}_{\me
-r+2},\Lq}_r + \bra{\pobr{\Lq,\var{H}{\Lq}}_r}_{\me 2r-2}.
\end{equation}
Otherwise a Dirac reduction is required.

Analogously, Lax operators \eqref{laxk2b} in the form
\begin{equation}
\Lq = u_N p^N + u_{N-1} p^{N-1} + ... + u_{3-r} p^{3-r} + u_{2-r}
p^{2-r}-q,
\end{equation}
form a proper submanifold for $r\me 2$ and  $N\me 2r-3$. Then,
the Poisson tensor has the form \eqref{tenk2a}. Otherwise a Dirac
reduction is required. The simplest case is for $r=1$ with
one-field reduction. Let
\begin{equation}
\overline{\Lq} =  \Lq + u = u_{N}p^{N} + u_{N-1}p^{N-1} + ... +
u_2 p^2 + u_1 p + u - q.
\end{equation}
The Dirac reduction with the constraint $u=0$ leads to the Poisson
map in the form
\begin{equation}\label{tenk2b}
\theta^{red}\bra{\var{H}{\Lq}} =
\bra{\pobr{\Lq,\var{H}{\Lq}}_1}_{\me 0} -
\pobr{\Dy{-1}res_1\pobr{\Lq,\var{H}{\Lq}}_1,\Lq}_1,
\end{equation}
which is generally nonlocal.

Hence, we know the Poisson structure for (2+1)-dispersionless
systems constructed from Poisson algebras, and since we are
interested in Hamiltonian systems, we shall now consider the
problem of their construction. The conserved quantities $H_i$ are
described by the Hamiltonian equations
\begin{equation}\label{Hameq}
\Lq_{t_i} = \theta d H_i(\Lq).
\end{equation}
First we have to find  cosymmetries (one-forms) $dH_i$ which are
gradients of Hamiltonians. Because we are using gradients of
Casimir functionals $dC_i$ to generate equations \eqref{laxh}, our
$dH_i$ are given by projections of $dC_i$ on subspaces spanned by
$dH_i$ in the form \eqref{dhk0},\eqref{dhk1} and \eqref{dhk2} for
$k=0,1$ and $2$, respectively. Then, we can apply the Lemma
\ref{fun} and hence Hamiltonians are defined as follows
\begin{equation}\label{Ham}
H_i(\Lq) = \int_0^1 Tr(dH_i(\lambda \Lq) \Lq) d\lambda =
\iint_{\Omega \times \Si} \int_0^1 res_r(dH_i(\lambda \Lq) \Lq)
d\lambda dx dy.
\end{equation}
For Lax operator $\Lq=\sum_{i=1}^n u_ip^i - q$ the gradients from
\eqref{Ham} are given by
\begin{equation}
dH_i(\lambda \Lq)=\sum_{i=1}^n \var{h}{(\lambda u_i)}(\lambda
u_1,\lambda u_2,...,\lambda u_n)p^{r-1-i}.
\end{equation}
Hence, by using definition of the residuum \eqref{res} we get that
\begin{equation}
res_r(dH_i(\lambda \Lq) \Lq) = \sum_{i=1}^n u_i \var{h}{(\lambda
u_i)}(\lambda u_1,\lambda u_2,...,\lambda u_n).
\end{equation}

Contrary to the (1+1)-dimensional case in the (2+1) case, the
functional densities contain terms with $x,y$ derivatives as well
as nonlocal terms. Nevertheless, all these additional terms appear
in a special form, namely they are expressed through the
pseudo-differential operators of the form $\Dq^k,\Dq^{-k}$ where
\begin{equation}\label{dq}
\Dq := \Dx{-1} \Dy{},\quad \Dq^{-1} := \Dy{-1} \Dx{}.
\end{equation}
Thus, the additional to \eqref{Ham} useful relation of calculation
of variations containing $\Dq$, derived from \eqref{grad}, is the
following one
\begin{equation}
\var{}{u} \iint_{\Omega \times \Si} f(u) \Dq^k g(u)\ dxdy =
\deriv{f(u)}{u} \Dq^k g(u) + \deriv{g(u)}{u} \Dq^k f(u).
\end{equation}

\section{A list of some (2+1)-dimensional dispersionless systems}

In this section we will display a list of the simplest nonlinear
dispersionless (2+1)-dimensional integrable systems. Calculating
the gradients $dC_n$ ($n$-highest order) given by \eqref{sol1} we
consider the Lax hierarchy
\begin{equation}\label{laxn}
\Lq_{t_{n}} = \pobr{\bra{dC_n}_{\me -r+k},\Lq}_r,\qquad n\in
\mathbb{Z}.
\end{equation}
The second hierarchy for $dC_n$ given by \eqref{sol2} can be
obtained by the transformation from Theorem \ref{rel}, which we
leave for the interested reader. We present the Hamiltonian
structure for particular choices of $r$. For $k=0$ and $k=1$ the
choice $n=1-r$ will always lead to the dynamics
$(u_i)_{t_{1-r}}=(1-r)(u_i)_x$ for the fields $u_i$ in $L$, so
that we may identify $t_{1-r}=\frac{1}{1-r}x$ in this cases. For
$(dC_n)_{\me -r+k}=L$ the equations become trivial, and then
$\Lq_{t_{n+r-k}}=\Lq_y$. For each choice of $k=0,1$ or $2$ and $N$
we will exhibit the first nontrivial of the nonlinear Lax
equations \eqref{laxn} associated with a chosen operator $\Lq$.

\paragraph{The case of $k=0$.}

\begin{example}
The dispersionless Kadomptsev-Petviashvili: $k=0, r=0, N=2$.

\rm The dispersionless Kadomptsev-Petviashvili (dKP) equation is a
(2+1)-dimensional extension of the dispersionless KdV equation.
The Lax operator for the (2+1)-dimensional dKP hierarchy has the
form
\begin{equation}
\Lq=p^2 + u - q.
\end{equation}
Then we derive for $(dC_3)_{\me 0} = p^3 + \frac{3}{2}up +
\frac{3}{4}\Dq u$
\begin{equation}
u_{t_3} = \frac{3}{2}uu_x + \frac{3}{4}\Dq u_y = \theta dH,
\end{equation}
where we get the Poisson tensor and the Hamiltonian
\begin{equation}
\theta = 2 \Dx{} ,\quad H = \frac{1}{8} \iint_{\Omega \times \Si}
(u^3 + \frac{3}{2} u\Dq^2u)\ dxdy.
\end{equation}
\end{example}

\begin{example}
The (2+1) Boussinesq hierarchy: $k=0, r=0, N=3$.

\rm The Lax operator is given by
\begin{equation}
\Lq= p^3 + up + v -q.
\end{equation}
We derive for $(dC_2)_{\me 0} = p^2 + \frac{2}{3} u$
\begin{equation}
\Matrix{t_2}{u}{v} = \Matrix{}{2v_x}{-\frac{2}{3}uu_x +
\frac{2}{3}u_y} = \theta dH.
\end{equation}
Eliminating the field $v$ from this equation we can derive the
(2+1)-dimensional 'dispersionless' Boussinesq equation
\begin{equation}
u_{tt} = \frac{4}{3}u_{xy} - \frac{2}{3}(u^2)_{xx}.
\end{equation}
The respective Poisson tensor and Hamiltonian are given in the
following form
\begin{equation}
\theta =
\pmatrix{cc}{
0& 3\Dx{}\\ 3\Dx{}&0
},\quad H = \frac{1}{3} \iint_{\Omega \times \Si} (-\frac{1}{9}u^3+v^2+\frac{1}{3}u\Dq u)\ dxdy.
\end{equation}
\end{example}

\begin{example}
The case: $k=0, r=0, N=4$.

\rm The Lax operator is
\begin{equation}
\Lq = p^4 + up^2 + vp + w -q,
\end{equation}
then for $(dC_2)_{\me 0} = p^2 + \frac{1}{2}u$ we have
\begin{equation}
\Matrixx{t_2}{u}{v}{w} =
\Matrixx{}{2v_x}{2w_x-uu_x}{\frac{1}{2}u_y-\frac{1}{2}u_xv} =
\theta dH,
\end{equation}
where
\begin{equation}
\theta =
\pmatrix{ccc}{
0& 0& 4\Dx{}\\
0& 4\Dx{}& 0\\
4\Dx{} &0 &\Dx{}u+u\Dx{}
},
\end{equation}
\begin{equation}
H = \frac{1}{4} \iint_{\Omega \times \Si}
(-\frac{1}{2}u^2v+2vw+\frac{1}{4}u\Dq u)\ dxdy.
\end{equation}
\end{example}

\paragraph{The case of $k=1$.}

\begin{example}
Three field hierarchy: $k=1, r\in \bb{Z}\setminus {2}$.

\rm The Lax operator has the form \eqref{laxk1a} with $N=2-r,\ m=r+1$
\begin{equation}\label{lax1Z}
\Lq = p^{2-r} + up^{1-r} + vp^{-r} + wp^{-r-1} - q.
\end{equation}
Then for $(dC_{2-r})_{\me -r+1} = p^{2-r} + up^{1-r}$ we have
\begin{equation}\label{lax1Ze}
\Matrixx{t_{2-r}}{u}{v}{w} =
\Matrixx{}{u_y+(2-r)v_x}{ru_xv+(1-r)uv_x+(2-r)w_x}{(1+r)u_xw+(1-r)uw_x}.
\end{equation}
This Lax operator forms a proper submanifold as regards the
condition $N\me 2r-1 \me -m$ only for $r=0,1$, otherwise a Dirac
reduction is required. Then for $r=0$
\begin{equation}
\Matrixx{t_2}{u}{v}{w} =
\Matrixx{}{u_y+2v_x}{uv_x+2w_x}{u_xw+uw_x} = \theta dH,
\end{equation}
where
\begin{equation}\label{tr0}
\theta=
\pmatrix{ccc}{
0& 0& 2\Dx{}\\ 0& 2\Dx{}& u\Dx{}-\Dy{}\\ 2\Dx{}& \Dx{}u-\Dy{}& 0
},
\end{equation}
\begin{align}
H= \frac{1}{16} \iint_{\Omega \times \Si} &\left ( 16vw - 2u^2\Dq v+8u\Dq w+\frac{1}{4}u^2\Dq u^2 \right .\nonumber\\
 &\quad \biggl .+4v\Dq v -u\Dq^2u^2+4u\Dq^2v+u\Dq^3u
\biggr )\ dxdy.
\end{align}
For $r=1$ we have
\begin{equation}\label{tr1}
\Matrixx{t_1}{u}{v}{w} = \Matrixx{}{u_y+v_x}{u_xv+w_x}{2u_xw} =
\theta dH,
\end{equation}
where
\begin{equation}
\theta=
\pmatrix{ccc}{
\Dy{}& \Dx{}v& 2\Dx{}w\\ v\Dx{}& \Dx{}w+w\Dx{}& 0\\ 2w\Dx{}& 0& 0
},\quad H= \frac{1}{2} \iint_{\Omega \times \Si} (u^2+2v)\ dxdy.
\end{equation}
\end{example}

\begin{example}
Dispersionless (2+1) Toda: $k=1, r\in \bb{Z}\setminus \{2\}$.

\rm The first admissible reduction $w=0$ of (\ref{lax1Z}) leads to the
two field Lax operator
\begin{equation}\label{lax1Zi}
\Lq = p^{2-r} + up^{1-r} + vp^{-r}-q.
\end{equation}
This Lax operator forms a proper submanifold only for $r=1$,
otherwise a Dirac reduction is required. Hence, for $r=1$ by
reduction $w=0$ \eqref{tr1} we get the (2+1)-dimensional
dispersionless Toda equation
\begin{equation}
\Matrix{t_1}{u}{v} = \Matrix{}{u_y+v_x}{u_xv} = \theta dH,
\end{equation}
where
\begin{equation}
\theta=
\pmatrix{cc}{
\Dy{}& \Dx{}v \\
v\Dx{}& 0
},\quad H= \frac{1}{2} \iint_{\Omega \times \Si} (u^2+2v)\ dxdy,
\end{equation}
known till now in a few non-Hamiltonian representations \cite{TT,KA,M-WWZ}.
Changing the independent coordinate $t'=t-y$ and eliminating $u$-field one
gets
\begin{equation}
(\ln v)_{tt'} = v_{xx}\quad \mbox{or} \quad \phi_{tt'} = ({\rm e}^{\phi_x})_x
\end{equation}
where $\phi_x = \ln v$. For $r=0$ we have
\begin{equation}
\Matrix{t_2}{u}{v} = \Matrix{}{u_y+2v_x}{uv_x},
\end{equation}
but we lose the Hamiltonian structure since the Poisson tensor
\eqref{tr0} is not reducible with the constraint $w=0$. Hence, the
Lax operator \eqref{lax1Zi} for $r=0$ generates equations which
are non-Hamiltonian.
\end{example}

The next admissible reduction $w=v=0$ of \eqref{lax1Ze} leads to
trivial equation $L_{t_{2-r}}=L_y$ since $(dC_{2-r})_{\me
-r+1}=L$.

\begin{example}
One field hierarchy: $k=1, r\in \bb{Z}\setminus \{2\}$.

\rm The Lax operator is given in the form
\begin{equation}\label{lax1Zii}
\Lq = p^{2-r} + (2-r)up^{1-r}-q.
\end{equation}
Then one finds for $\bra{dC_{3-r}}_{\me -r+1} = p^{3-r} + (3-r)up^{2-r} +
\bra{\frac{3-r}{2-r}\Dq u+\frac{1}{2}(3-r)u^2}p^{1-r}$ a whole family of
(2+1)-dimensional dispersionless one-field systems
\begin{equation}
u_{t_{3-r}} = -\frac{1}{2} (3-r)(1-r)u^2u_x+\frac{r(3-r)}{2-r} uu_y+\frac{3-r}{(2-r)^2} \Dq u_y+\frac{(3-r)(1-r)}{2-r} u_x\Dq u,
\end{equation}
derived for the first time in \cite{Bl2}, including the modified
dKP as a special case of $r=0$. This Lax operator forms a proper
submanifold only for $r=1$, in other cases a Dirac reduction is
required. For $r=1$ we get
\begin{equation}
u_{t_2} = 2uu_y+2\Dq u_y = \theta dH,
\end{equation}
where
\begin{equation}
\theta = \Dy{},\quad H= \iint_{\Omega \times \Si}
(\frac{1}{3}u^3+u\Dq u)\ dxdy.
\end{equation}
For $r=0$ we get
\begin{equation}
u_{t_3} = -\frac{3}{2} u^2u_x+\frac{3}{4} \Dq u_y+\frac{3}{2}
u_x\Dq u,
\end{equation}
and by Dirac reduction of \eqref{tr0} with the constraint $w=v=0$
we get the formal Poisson tensor
\begin{equation}
\theta^{red} = 8\Dx{} (\Dy{}-2u\Dx{})^{-1} \Dx{} (\Dy{}-2\Dx{}u)^{-1}
\Dx{},
\end{equation}
and the related sympletic tensor
\begin{equation}
J=\bra{\theta^{red}}^{-1}= \frac{1}{8} (\Dq -2u)\Dx{-1}(\Dq -2u),
\end{equation}
such that $Ju_{t_3}=dH$, where
\begin{align}
H = \frac{3}{32} \iint_{\Omega \times \Si} &\left ( -\frac{1}{3}u^6+u\Dq u^4+\frac{1}{2}u^2\Dq^2u^2-u^2(\Dq u)^2\right.\nonumber\\
 &\quad \left .+\frac{1}{3}(\Dq u)^3 -u\Dq^3u^2+\frac{1}{2}u\Dq^4u
\right)\ dxdy.
\end{align}
\end{example}

\begin{example}
Three field hierarchy: $k=1, r\in \bb{Z}$.

\rm This case does not exist in (1+1)-dimension. The Lax operator has
the form \eqref{laxk1b} with $\ m=r+2$
\begin{equation}\label{lax1p}
\Lq = up^{-r} + vp^{-r-1} + wp^{-r-2} - q.
\end{equation}
Then for $(dC_{2-r})_{\me -r+1} = p^{2-r} + (r-2)\Dq^{-1}u
p^{1-r}$ we have
\begin{equation}
\Matrixx{t_{2-r}}{u}{v}{w} =
(r-2)\Matrixx{}{ru\Dq^{-1}u_x+(1-r)u_x\Dq^{-1}u-v_x}{(1+r)v\Dq^{-1}u_x+(1-r)v_x\Dq^{-1}u-w_x}{(2+r)w\Dq^{-1}u_x+(1-r)w_x\Dq^{-1}u}.
\end{equation}
This Lax operator forms a proper submanifold as regards the
condition $2r-1 \me -m$ only for $r=0$, otherwise a Dirac
reduction is required. Then for $r=0$
\begin{equation}\label{tpr0}
\Matrixx{t_2}{u}{v}{w} =
-2\Matrixx{}{u_x\Dq^{-1}u-v_x}{v\Dq^{-1}u_x+v_x\Dq^{-1}u-w_x}{2w\Dq^{-1}u_x+w_x\Dq^{-1}u}
= \theta dH,
\end{equation}
where
\begin{equation}
\theta=
\pmatrix{ccc}{
0& -\Dy{}& 0\\ -\Dy{}& 0& 0\\ 0& 0& \Dx{}w+w\Dx{}
},
\end{equation}
and
\begin{eqnarray}
H= \iint_{\Omega \times \Si}
\left (-2u\Dq^{-1}w-v\Dq^{-1}v+v\bra{\Dq^{-1}u}^2\right )\ dxdy.
\end{eqnarray}
For $r=1$ we have
\begin{equation}
\Matrixx{t_1}{u}{v}{w} =
-\Matrixx{}{u\Dq^{-1}u_x-v_x}{2v\Dq^{-1}u_x-w_x}{3w\Dq^{-1}u_x} = \theta
dH.
\end{equation}
We derive the Poisson tensor from \eqref{tenk1b}, then
\begin{equation}\label{tpr1}
 \theta^{red} =
 \pmatrix{ccc}{
-u\Dq^{-1}\Dx{}u+\Dx{}v+v\Dx{}& -2u\Dq^{-1}\Dx{}v+2\Dx{}w+w\Dx{}& -3u\Dq^{-1}\Dx{}w\\
-2v\Dq^{-1}\Dx{}u+\Dx{}w+2w\Dx{}& -4v\Dq^{-1}\Dx{}v& -6v\Dq^{-1}\Dx{}w\\
-3w\Dq^{-1}\Dx{}u& -6w\Dq^{-1}\Dx{}v& -9w\Dq^{-1}\Dx{}w
},
\end{equation}
and
\begin{equation}
H = \iint_{\Omega \times \Si} u\ dxdy.
\end{equation}
\end{example}

\begin{example}
Two field hierarchy: $k=1, r\in \bb{Z}$.

\rm The first admissible reduction $w=0$ of (\ref{lax1p}) leads to the
two field Lax operator
\begin{equation}
\Lq = up^{-r} + vp^{-r-1}-q.
\end{equation}
This Lax operator forms a proper submanifold only for $r=0$,
otherwise a Dirac reduction is required. Hence, for $r=0$ by
reduction $w=0$ of \eqref{tpr0} we get
\begin{equation}
\Matrix{t_2}{u}{v} =
-2\Matrix{}{u_x\Dq^{-1}u-v_x}{v\Dq^{-1}u_x+v_x\Dq^{-1}u} = \theta
dH,
\end{equation}
where
\begin{equation}\label{hamr0}
\theta=
\pmatrix{cc}{
0& -\Dy{} \\
-\Dy{}& 0
},
\end{equation}
\begin{equation}
H= \frac{2}{3} \iint_{\Omega \times \Si}
\bra{-v\Dq^{-1}v+v\bra{\Dq^{-1}u}^2}
\ dxdy.
\end{equation}
For $r=1$ we have
\begin{equation}
\Matrix{t_1}{u}{v} = -\Matrix{}{u\Dq^{-1}u_x-v_x}{2v\Dq^{-1}u_x} =
\theta dH.
\end{equation}
We derive the Poisson tensor from \eqref{tpr1} with the constraint
$w=0$, then
\begin{equation}\label{tpr1a}
\theta^{red} =
\pmatrix{cc}{
-u\Dq^{-1}\Dx{}u+\Dx{}v+v\Dx{}& -2u\Dq^{-1}\Dx{}v\\
-2v\Dq^{-1}\Dx{}u& -4v\Dq^{-1}\Dx{}v
},
\end{equation}
and
\begin{equation}
H = \iint_{\Omega \times \Si} u\ dxdy.
\end{equation}
\end{example}

\begin{example}
One field hierarchy: $k=1, r\in \bb{Z}$.

\rm The second admissible reduction $w=v=0$ of (\ref{lax1p}) leads to
the one field Lax operator
\begin{equation}
\Lq = up^{-r} -q.
\end{equation}
This Lax operator does not form a proper submanifold as the
condition $2r-1\me -m$ is violated, hence a Dirac reduction is
required. For $r=0$ by reduction $v=w=0$ of \eqref{tpr0} we get
\begin{equation}
u_{t_2} = -2u_x\Dq^{-1}u,
\end{equation}
but we lose the Hamiltonian structure as the Poisson tensor
\eqref{hamr0} is not Dirac reducible with constraint $v=w=0$. For
$r=1$ we have
\begin{equation}
u_{t_1} = -u\Dq^{-1}u_x = \theta dH.
\end{equation}
We derive the Poisson tensor from \eqref{tpr1a} with the
constraint $w=0$, then
\begin{equation}
\theta^{red} = -u\Dq^{-1}\Dx{}u,
\end{equation}
and
\begin{equation}
H = \iint_{\Omega \times \Si} u\ dxdy.
\end{equation}
\end{example}

\paragraph{The case of $k=2$.}

\begin{example}
One field hierarchy: $k=2, r\in \bb{Z}\setminus \{2\}$.

\rm The simplest admissible Lax operator is given by
\begin{equation}
\Lq = u^{2-r}p^{2-r}-q.
\end{equation}
This case does not exist in (1+1)-dimension. In this case we have
to consider separately two cases: $r\neq 1$ and $r=1$. Then, one
finds again a whole family of (2+1)-dimensional dispersionless
one-field systems \cite{Bl2} including a dispersionless
(2+1)-dimensional Harry Dym equation as a special case of $r=0$:
\paragraph{$r\neq 1:$}
for $\bra{dC_{3-r}}_{\me -r+2} = u^{3-r}p^{3-r} +
\frac{3-r}{(r-1)(2-r)}u^{2-r}\Dq u^{r-1}p^{2-r}\ $ one finds
\begin{equation}
u_{t_{3-r}} = \frac{3-r}{(r-1)(2-r)}u_y\Dq
u^{r-1}+\frac{3-r}{(2-r)^2} u\Dq u^{r-2}u_y,
\end{equation}
\paragraph{$r=1:$}
for $\bra{dC_{2}}_{\me 1} = u^2p^2 + 2u\Dq\bra{\ln u}p\ $ one finds
\begin{equation}\label{ofhr1}
u_{t_{2}} = 2u_y\Dq \ln u+ 2u\Dq \bra{\ln u}_y.
\end{equation}
To get $\theta$, we have to make a Dirac reduction as the
conditions $r\me 2,N\me 2r-3$ are violated. Poisson tensor for
$r=1$ is given by \eqref{tenk2b}, then we get for \eqref{ofhr1}
the Hamiltonian structure, where
\begin{equation}
\theta^{red} = u\Dq^{-1}\Dx{}u,\quad H= \iint_{\Omega \times \Si}
\bra{\ln u\Dq^3\ln u+\frac{1}{3}\bra{\Dq \ln u}^3}\ dxdy.
\end{equation}
\end{example}

\begin{example}
Two field hierarchy: $k=2, r\in \bb{Z}\setminus \{3\}$.

\rm The Lax operator is given by
\begin{equation}
L = up^{3-r}+vp^{2-r}-q.
\end{equation}
This case is nonreducible to (1+1)-dimension. Then, one finds for
$(dC_{2-r})_{\me-r+2}=u^{\frac{2-r}{3-r}}p^{2-r}$
\begin{equation}
\Matrix{t_{2-r}}{u}{v} = \frac{2-r}{3-r} u^{\frac{-1}{3-r}}
\Matrix{}{(3-r)uv_x-(2-r)u_xv}{u_y}.
\end{equation}
To get $\theta$ we have to make a Dirac reduction as the
conditions $r\me 2,N\me 2r-3$ are violated. The Poisson tensor for
$r=1$ is given by \eqref{tenk2b}, then
\begin{equation}
\Matrix{t_1}{u}{v} = \frac{\sqrt{u}}{2u}
\Matrix{}{2uv_x-u_xv}{u_y} = \theta^{red} dH,
\end{equation}
where
\begin{equation}
\theta^{red}dH =
\pmatrix{cc}{
4u\Dq^{-1}\Dx{}u& 2u\Dq^{-1}\Dx{}v\\
2v\Dq^{-1}\Dx{}u& 2v\Dq^{-1}\Dx{}v+\Dx{}u+u\Dx{}
},
\end{equation}
and
\begin{equation}
H= \iint_{\Omega \times \Si} \bra{
-\frac{1}{8}\frac{v^3\sqrt{u}}{u^2}-\frac{1}{3}u\Dq
\frac{v\sqrt{u}}{u}+\frac{3}{4} \frac{v\sqrt{u}}{u}\Dq \ln u }\
dxdy.
\end{equation}
\end{example}

\begin{example}
Two field hierarchy: $k=2, r\in \bb{Z}\setminus \{2\}$.

\rm The Lax operator is given by
\begin{equation}
\Lq = u^{2-r}p^{2-r}+vp^{1-r}+p^{-r}-q.
\end{equation}
Then, for $(dC_{2-r})_{\me-r+2}=u^{2-r}p^{2-r}$ one finds
\begin{equation}
\Matrix{t_{2-r}}{u}{v} =
\Matrix{}{u_y-(1-r)u_xv+uv_x}{(2-r)ru^{1-r}u_x}.
\end{equation}
This Lax operator forms a proper submanifold only for $r=1$,
otherwise a Dirac reduction is required. Hence, for $r=1$ we get
\begin{equation}
\Matrix{t_1}{u}{v} = \Matrix{}{u_y+uv_x}{u_x} = \theta dH,
\end{equation}
where
\begin{equation}
\theta dH =
\pmatrix{cc}{
0& u\Dx{}\\ \Dx{}u& -\Dy{}
},
\end{equation}
and
\begin{equation}
H= \iint_{\Omega \times \Si} \bra{ u+\frac{1}{2}v^2+v\Dq \ln
u+\frac{1}{2}\ln u\ \Dq^2\ln u }\ dxdy.
\end{equation}
\end{example}

\begin{example}
Three field hierarchy: $k=2, r\in \bb{Z}$.

\rm The Lax operator is given by
\begin{equation}
\Lq = up^{2-r}+vp^{1-r}+wp^{-r}+p^{-r-1}-q.
\end{equation}
Then for $(dC_{2-r})_{\me-r+2}=up^{2-r}$ one finds
\begin{equation}
\Matrixx{t_{2-r}}{u}{v}{w} =
\Matrixx{}{u_y-(1-r)u_xv+(2-r)uv_x}{ru_xw+(2-r)uw_x}{(1+r)u_x}.
\end{equation}
This Lax operator forms a proper submanifold only for $r=1$,
otherwise a Dirac reduction is required. Hence, for $r=1$ we get
\begin{equation}
\Matrixx{t_1}{u}{v}{w} =
\Matrixx{}{u_y+uv_x}{u_xw+uw_x}{2u_x}=\theta dH,
\end{equation}
where
\begin{equation}
\theta =
\pmatrix{ccc}{
0& u\Dx{}& 0\\ \Dx{}u& -\Dy{}& 0\\ 0& 0& 2\Dx{}
},
\end{equation}
and
\begin{equation}
H= \iint_{\Omega \times \Si} \bra{ \frac{1}{2}v^2+uw+v\Dq \ln
u+\frac{1}{2}\ln u\ \Dq^2 \ln u }\ dxdy.
\end{equation}
\end{example}

\end{document}